\documentclass{iopart}
\usepackage{latexsym}
\usepackage{amsfonts}
\usepackage{amsbsy}
\usepackage{mathrsfs}
\usepackage{iopams}
\newtheorem{theorem}{Theorem}

\begin{document}
\title[Time boundary terms and Dirac constraints]{Time boundary terms and
Dirac constraints}
\author{Alejandro Gallardo}

\address{Departamento de F\'{\i}sica, Centro de Investigaci\'on y de Estudios
Avanzados del Instituto Polit\'ecnico Nacional, Avenida Instituto
Polit\'ecnico Nacional 2508, San Pedro Zacatenco, 07360, Gustavo A. Madero,
Ciudad de M\'exico, M\'exico.}

\eads{\mailto{agallardo@fis.cinvestav.mx}}

\begin{abstract}
Time boundary terms usually added to action principles are
systematically handled in the framework of Dirac's canonical
analysis. The procedure begins with the introduction of the boundary
term into the integral Hamiltonian action and then the resulting
action is interpreted as a Lagrangian one to which Dirac's method is
applied. Once the general theory is developed, the current procedure
is implemented and illustrated in various examples which are
originally endowed with different types of constraints.
\end{abstract}

\pacs{45.20.Jj, 45.20.-d, 45.05.+x, 03.65.Ca}


\maketitle

\section{Introduction}
Time boundary terms are frequently introduced in action principles for both
Lagrangian and Hamiltonian systems. On one hand, boundary terms are needed to
select a complete set of commuting variables which are going to be fixed at
the time boundary and thus, the choice of one of these sets is a motivation
for choosing a particular boundary term or another \cite{HV1992,HTV1992}.

On the other hand, in gauge theories with a finite number of degrees of
freedom, it is usual to deal with Hamiltonian action principles which are not
fully gauge-invariant under the gauge transformation generated by their
corresponding first-class constraints and in these cases it is possible to add
boundary terms to these actions in such a way that the resulting action is
fully gauge-invariant under the gauge transformation generated by the
first-class constraints involved \cite{HV1992,HTV1992,T1994,MV2002,cmv}. This
is also the case for field theory \cite{MV2001,MV2003}.

Independently of the motivation at hand to introduce boundary terms, it would
be useful to have a general formalism or a recipe to handle them in the
theoretical framework of the canonical analysis. The ideas developed in this
paper are along this way of thinking. The procedure, proposed in Ref.
\cite{MV2002} some years ago but developed here for the first time, consists
in first to introduce the boundary term into the integral action and then to
interpret the resulting action as a Lagrangian one to which the canonical
analysis can be applied. Such a procedure has the disadvantage of enlarging
the original set of variables but has the advantage of being completely
systematic to handle boundary terms.

\section{Theoretical framework}
The starting point is a Hamiltonian system described by an action principle of
the form \cite{D1,Dbook,Hbook}
\begin{equation}\label{star-act}
S[q^i,p_i,u^a,v^\alpha]=\int_{\tau_1}^{\tau_2}d\tau\left[\dot{q}^ip_i-H_E\right],\qquad
i=1,...,N,
\end{equation}
where $H_E=H_0+u^a\gamma_a+v^{\alpha}\varphi_\alpha$ is the extended
Hamiltonian, $\gamma_a$ are first-class constraints $(a=1,...,A)$,
$\varphi_\alpha$ are second-class constraints
($\alpha=1,...,\mathcal{A}$), and $H_0$ is the first-class canonical
Hamiltonian; $u$'s and $v$'s are their respective Lagrange
multipliers. This system has $\frac{1}{2} \left ( 2N-\mathcal{A}-2A
\right )$ degrees of freedom. Following the ideas mentioned in the
Introduction, a time boundary term is added to the action
(\ref{star-act})
\begin{equation}\label{boundterm}
S[q^i,p_i,u^a,v^\alpha] =
\int_{\tau_1}^{\tau_2}d\tau\left[\dot{q}^ip_i-H_E\right] - B(q^i,p_i)
\mid^{\tau_2}_{\tau_1}.
\end{equation}
Notice that, by hypothesis, $B$ is explicitly $\tau$-independent. The
introduction of the time boundary term into the action principle yields to
\begin{eqnarray}\label{act-bound}
S[q^i,p_i,u^a,v^\alpha] &=& \int_{\tau_1}^{\tau_2}d \tau \left[
\dot{q}^i p_i - H_E - \frac{d}{d\tau}B \right] \nonumber\\
&=& \int_{\tau_1}^{\tau_2} d\tau\left[ \dot{q}^i p_i-H_E-\frac{\partial
B}{\partial q^i}\dot{q}^i-\frac{\partial B}{\partial p_i} \dot{p}_i \right].
\end{eqnarray}
The next step is to interpret the action (\ref{act-bound}) as a
Lagrangian one and so to define the momenta
$(\pi_{x^{\mu}})=(\pi_{q^i}, \pi_{p_i}, \pi_{u^a},
\pi_{v^{\alpha}})$ canonically conjugated to the configuration
variables $(x^\mu)=(q^i,p^i,u^a,v^{\alpha})$, which leads to the
following primary constraints
\begin{eqnarray}\label{const:1}
\phi_{q^i} &:=& \pi_{q^i}-p_i+\frac{\partial B}{\partial
q^i}\approx0,\qquad\,\,\,  \phi_{u^a}:=\pi_{u^a}\approx0,\nonumber\\
\phi_{p^i} &:=& \pi_{p^i}+\frac{\partial B}{\partial
p^i}\approx0,\qquad\qquad\phi_{v^\alpha}:=\pi_{v^\alpha} \approx 0.
\end{eqnarray}

Performing the Legendre transformation, the canonical Hamiltonian $H_c$ is
computed
\begin{eqnarray}
\fl H_c &=& \pi_{q^i} {\dot q}^i + \pi_{p^i}{\dot p}^i + \pi_{u^a}
{\dot u}^a+\pi_{v^{\alpha}}{\dot v}^{\alpha}-\left(\dot{q}^ip^i
-H_E-\frac{\partial B}{\partial q^i}\dot{q}^i-\frac{\partial
B}{\partial p^i}\dot{p}^i\right) \nonumber\\
\fl &=& H_E.
\end{eqnarray}
Therefore, the action principle is promoted to have the following Hamiltonian
form
\begin{equation}\label{act-bod}
S[x^\mu,\pi_{x^{\mu}},\lambda^{\mu}]:=
\int_{\tau_1}^{\tau_2}d\tau\left[\dot{x}^\mu \pi_{x^{\mu}}- H_c
-\lambda^{\mu}\phi_{\mu}\right],
\end{equation}
where $\lambda^{\mu}$ are the corresponding Lagrange multipliers. From the
variation of the action of Eq. (\ref{act-bod}), the dynamical equations
\begin{eqnarray}\label{eqmo:1}
\dot{\pi}_{q^i} &=&-\frac{\partial H_c}{\partial q^i}- \lambda^{q^j}
\frac{\partial\phi_{q^j}}{\partial q^i}-
\lambda^{p^j}\frac{\partial\phi_{p^j}}{\partial q^i}, \nonumber\\
\dot{\pi}_{p^i} &=&-\frac{\partial H_c}{\partial p^i}- \lambda^{q^j}
\frac{\partial\phi_{q^j}}{\partial p^i}-
\lambda^{p^j}\frac{\partial\phi_{p^j}}{\partial p^i},\nonumber\\
\dot{\pi}_{u^a} &=&-\gamma_a, \nonumber\\
\dot{\pi}_{v^\alpha} &=&-\varphi_\alpha, \nonumber\\
\dot{x}^\mu &=& \lambda^{\mu},
\end{eqnarray}
together with the constraints (\ref{const:1}) are obtained. By using
the equations of motion, the evolution of the primary constraints
$\phi_{q^i}$ and $\phi_{p^i}$ is computed and it is easy to check
that the Lagrange multipliers associated to these constraints get
fixed
\begin{equation}\label{mult:1}
\lambda^{q^i} =\frac{\partial H_E}{\partial
p^i},\qquad\qquad\lambda^{p^i}=-\frac{\partial H_E}{\partial q^i},
\end{equation}
while the time evolution of $\phi_{u^a}$ and $\phi_{v^\alpha}$
produces the following secondary constraints
\begin{eqnarray}
\gamma_a\approx0,\qquad\qquad\varphi_\alpha\approx0.
\end{eqnarray}
These constraints are the first- and second-class constraints of the
original system. Since the evolution of the secondary constraints
gives us relations among the Lagrange multipliers $\lambda^{q^i}$
and $\lambda^{p^i}$ which strongly vanish after plugging into them
the explicit form for the Lagrange multipliers given in Eq.
(\ref{mult:1}), the process ends and no more constraints arise.

In order to classify the complete set of constraints
$(\phi_I)=(\phi_{q^i},\phi_{p^i},\phi_{u^a},\phi_{v^\alpha},\gamma_a,\varphi_\alpha)$,
their Poisson brackets are computed and expressed in matrix form,
namely,
\begin{eqnarray}\label{matrix}
(\{\phi_I,\phi_J\})=\left(
\begin{array}{cccccc}
{\bf 0}            & -\delta_{ij} & {\bf 0} & {\bf 0} &-\frac{\partial\gamma_b}{\partial q^i}&-\frac{\partial\varphi_\beta}{\partial q^i} \\
\delta_{ij}  & {\bf 0}            & {\bf 0} & {\bf 0} &-\frac{\partial\gamma_b}{\partial p^i}&-\frac{\partial\varphi_\beta}{\partial p^i} \\
{\bf 0}            & {\bf 0}            & {\bf 0} & {\bf 0} & {\bf 0} & {\bf 0} \\
{\bf 0}            & {\bf 0}            & {\bf 0} & {\bf 0} & {\bf 0} & {\bf 0} \\
\frac{\partial\gamma_a}{\partial q^j}      &\frac{\partial\gamma_a}{\partial p^j}      & {\bf 0} & {\bf 0} & {\bf 0} & {\bf 0} \\
\frac{\partial\varphi_\alpha}{\partial q^j}&\frac{\partial\varphi_\alpha}{\partial p^j}& {\bf 0} & {\bf 0} & {\bf 0} & {\bf 0} \\
\end{array}
\right),
\end{eqnarray}
where $I,J=1,...,2(N+A+\mathcal{A})$. This matrix has a vanishing
determinant which tells us that there is at least one first-class
constraint. It is easy to check that the matrix (\ref{matrix}) has
$(2A+\mathcal{A})$ null vectors which implies that the constraints
must be redefined. By using these null vectors, we build an inverted
matrix $M_I{}^J$ in order to define an equivalent set of
constraints, i.e, $\tilde{\phi}_I:=M_I{}^J\phi_J$ \cite{Hbook},
namely
\begin{eqnarray}
\fl \widetilde{\phi}_I:=\left(\begin{array}{c}
  \phi_{u^a} \\
  \phi_{v^\alpha} \\
  \delta_a \\
  \phi_{q^i} \\
  \phi_{p^i} \\
  \varphi_\alpha
\end{array}\right)=\left(
               \begin{array}{cccccc}
               {\bf 0} & {\bf 0} & \delta_{ab} & {\bf 0} & {\bf 0} & {\bf 0} \\
               {\bf 0} & {\bf 0} & {\bf 0}  & \delta_{\alpha\beta}& {\bf 0} & {\bf 0} \\
               \frac{\partial\gamma_a}{\partial p^j} &-\frac{\partial\gamma_a}{\partial q^j} & {\bf 0} & {\bf 0}  & \delta_{ab} & {\bf 0} \\
               \delta_{ij} &{\bf 0} & {\bf 0} & {\bf 0} &  {\bf 0} & {\bf 0} \\
               {\bf 0} & \delta_{ij} & {\bf 0} & {\bf 0} & {\bf 0} & {\bf 0} \\
               {\bf 0} & {\bf 0} & {\bf 0} & {\bf 0} & {\bf 0} & \delta_{\alpha\beta} \\
               \end{array}
\right)\left(
         \begin{array}{c}
           \phi_{q^j} \\
           \phi_{p^j} \\
           \phi_{u^b} \\
            \phi_{v^\beta} \\
           \gamma_b \\
           \varphi_\beta \\
         \end{array}
       \right),
\end{eqnarray}
where $\det(M_I{}^J)=1$. With this equivalent set of constraints the
Poisson brackets among the constraints $\tilde{\phi}_I$ are computed
again which implies that
\begin{eqnarray}\label{fcc-g}
\phi_{u^a}:=\pi_{u^a}\approx0,\qquad
\phi_{v^\alpha}:=\pi_{v^\alpha}\approx0,\nonumber\\
\delta_{a}:=\gamma_a+\frac{\partial\gamma_a}{\partial
p^i}\phi_{q^i}-\frac{\partial\gamma_a}{\partial
q^i}\phi_{p^i}\approx 0,
\end{eqnarray}
are $e=(2A+\mathcal{A})$ first-class constraints, $(\Gamma_e)$, and
that
\begin{eqnarray}\label{scc-g}
\fl \phi_{q^i}:=\pi_{q^i}-p_i+\frac{\partial B}{\partial q^i}\approx
0, \quad \phi_{p^i} := \pi_{p^i}+\frac{\partial B}{\partial
p^i}\approx 0, \quad \varphi_\alpha\approx 0,
\end{eqnarray}
are  $\xi=(2N+\mathcal{A})$ second-class constraints, $(\chi_\xi)$.

At this point, it is worth noticing that the second-class
constraints $\varphi_\alpha$ arise from the time evolution of the
first-class constraints $\phi_{v^\alpha}$. From this fact, we
observe that the system analyzed here belongs to that ones which ``
cross the class-line in the constraint algorithm " as was pointed
out in \cite{gotay}.\\

Following Dirac's method, the first-class Hamiltonian must be built,
which can be achieved by plugging into $H= H_c
+\lambda^q_{exp}\chi_q$ the explicit form for the Lagrange
multipliers $\lambda^q_{exp}$ given in Eq. (\ref{mult:1}). Thus the
extended action principle becomes
\begin{equation}\label{act:1}
S[x^\mu,\pi_{x^\mu},\lambda^e,\Lambda^\xi]:=\int_{\tau_1}^{\tau_2}d\tau[\dot{x}^\mu
\pi_{x^\mu}-H-\lambda^e\Gamma_e-\Lambda^\xi\chi_\xi].
\end{equation}

Finally, the knowledge of the type of the constraints allows us the
computation of the degrees of freedom for the system, which turns
out to be $\frac{1}{2}[2N-\mathcal{A}-2A]$ and it agrees with the
original counting.

As was pointed out at the introduction one motivation in order to
add boundary terms is to built action principles which are fully
invariant under the gauge transformation generated by the
first-class constraints, therefore we have to analyze the boundary
term which arises from these transformations, such term is given by
\begin{eqnarray}
 M=\varepsilon^e\left(\frac{\partial \Gamma_e}{\partial
\pi_{x^\mu}}\pi_{x^\mu}-\Gamma_e\right).
\end{eqnarray}
where $\varepsilon^e$ are $(2A+\mathcal{A})$ gauge parameters. From
the functional form of the constraints $\phi_{u^a}$ and
$\phi_{v^\alpha}$, it is easy to see that they are linear and
homogeneous in the momenta $\pi_\mu$ so their corresponding boundary
term turns out to be zero. Therefore the unique contribution to the
boundary term $M$ is due to $\delta_a$, i.e.,
\begin{eqnarray}
M_{\delta}= \varepsilon^a\left(\frac{\partial \gamma_a}{\partial
p_i}p_i-\gamma_a +\frac{\partial \gamma_a}{\partial
q^i}\frac{\partial B}{\partial p_i}- \frac{\partial
\gamma_a}{\partial p_i}\frac{\partial B}{\partial q^i}\right).
\end{eqnarray}
Therefore, the extended action could be fully gauge invariant or not
depending on which boundary term, B, had been chosen
\cite{MV2002,cmv}.

In summary, the introduction of a time boundary term into action principle
(\ref{star-act}) and the application of the Dirac's method implies an
enlargement of the phase space and the appearance of second-class constraints
which contain the information about the time boundary.

\section{Examples}
The general theory is implemented in the following examples.

\subsection{Harmonic oscillator without time boundary term}
This example is relevant because it illustrates just one part of the whole
procedure developed in last section: the interpretation of the original
Hamiltonian action principle as a Lagrangian one. In other words, there is not
a boundary term at the time boundary.

Thus, the analysis begins with the following Hamiltonian action
principle for the non-relativistic one-dimensional harmonic
oscillator
\begin{eqnarray}\label{paraiho}
 S[q,p] &:=&
\int^{\tau_2}_{\tau_1}d\tau\left[p\dot{q}-\frac{p^2}{2m}
-\frac{1}{2}m\omega^2q^2\right],
\end{eqnarray}
The system has one physical degree of freedom. Following the
framework described in last section, the integrand is interpreted as
a Lagrangian and so from the definition of the momenta
$(\pi_{x^{\mu}})=(\pi_q,\pi_p)$ canonically conjugate to the
coordinates $(x^{\mu})=(q,p)$, the primary constraints
\begin{eqnarray}\label{pcho}
\phi_q=\pi_q-p \approx 0,\quad\mbox{and}\quad\phi_p =\pi_p\approx0
\end{eqnarray}
arise. With this information, the canonical Hamiltonian is computed
and it turns out to be $ H_c=\frac{p^2}{2m}
+\frac{1}{2}m\omega^2q^2$. Thus, the total action principle becomes
\begin{eqnarray}
S[x^\mu,\pi_{x^\mu},\lambda^\mu]=\int_{\tau_1}^{\tau_2}d\tau
\left[\dot{x}^\mu\pi_{x^\mu}-H_c-\lambda^\mu\phi_\mu\right].
\end{eqnarray}
By consistency, the evolution of the constraints (\ref{pcho}) is computed
which implies that some of the Lagrange multipliers get fixed
\begin{eqnarray}\label{mult:3}
\lambda^q=\frac{p}{m}\quad\mbox{and}\quad\lambda^p=-um\omega^2q,
\end{eqnarray}
Therefore, all the Lagrange multipliers has been fixed and so there
are no more constraint. A straightforward computation shows that the
Poisson brackets between the constraints are
$\{\phi_i,\phi_j\}=-\delta_{ij}$ where $i,j=1,2$ and hence $(\phi_q,
\phi_p)$ are second-class constraints. The first-class canonical
Hamiltonian becomes $H=\frac{p^2}{2m}-\frac{1}{2}m\omega^2q^2
+\frac{\pi_qp}{m}-m\omega^2q\pi_p$. The counting of the number of
degrees of freedom is $\frac{1}{2} (2\times 2-2)=1$, which is in
agreement with the original description of the system given by the
action (\ref{paraiho}).

Finally, due to there are no first-class constraints the boundary
term $M$ will not exist.

\subsection{Parameterized harmonic oscillator with time boundary term}

The next example is the parameterized non-relativistic one-dimensional harmonic
oscillator. The example is relevant because on it is seen the role of the
boundary term in the gauge invariance of the action. As starting point, we
take the Hamiltonian action principle for the parameterized non-relativistic
one-dimensional harmonic,
\begin{eqnarray}\label{act-ob}
S[q,t,p_q,p_t;u]:=\int^{\tau_2}_{\tau_1}d\tau\left[p\dot{q}+p_t\dot{t}-H_E\right],
\end{eqnarray}
where the extended Hamiltonian, $H_E=u\gamma$, is composed by the
first-class constraint $\gamma:=p_t+\frac{1}{2m}
\left(p^2+m^2\omega^2q^2\right)$ and the Lagrange multiplier $u$. On
the other hand, we take the boundary term
$B=B(q,t,p,p_t)|_{\tau_1}^{\tau_2}$. By applying the method, we add
this boundary term to the action principle (\ref{act-ob}), namely
\begin{eqnarray}\label{a+b}
S[q,t,p_q,p_t,u]=\int_{\tau_1}^{\tau_2}d\tau\left[\dot{q}p+\dot{t}p_t-H_E-
\frac{dB}{d\tau}\right].
\end{eqnarray}
As next step, the integrand of (\ref{a+b}) is interpreted as
Lagrangian ones, i.e.,
$\mathcal{L}:=\dot{q}p+\dot{t}p_t-u\left[p_t+\frac{1}{2m}
\left(p^2+m^2\omega^2q^2\right)\right]-\frac{\partial B}{\partial
q}\dot{q}-\frac{\partial B}{\partial t}\dot{t}- \frac{\partial
B}{\partial p}\dot{p}-\frac{\partial B}{\partial p_t}\dot{p}_t$. The
canonical analysis starts by the definition of the momenta,
$\pi_{x^\mu}=(\pi_q,\pi_t,\pi_p,\pi_{p_t},\pi_u)$, canonically
conjugated to the coordinates, $x^\mu=(q,t,p_q,p_t,u)$. From these
definitions it is easy to see that five primary constraints arise
\begin{eqnarray}
\begin{array}{ccccccc}
\phi_q&:=&\pi_q-p+\frac{\partial B}{\partial q}\approx0,      &
&\phi_t&:=& \pi_t-p_t+
\frac{\partial B}{\partial t}\approx0,\\
\phi_{p_t}&:=&\pi_{p_t}+\frac{\partial B}{\partial
p_t}\approx0,\quad\quad & &\phi_{p}&:=&\pi_{p}
+\frac{\partial B}{\partial p}\approx0,\quad\quad\\
\phi_{u}&:=&\pi_{u}\approx0.\qquad\qquad\\
\end{array}
\end{eqnarray}
Using this information, the canonical Hamiltonian becomes
$H_c=u\left[p_t+\frac{1}{2m}\left(p^2+m^2\omega^2q^2\right)\right]$
and the total action principle acquires the form
\begin{eqnarray}
\fl S[x^\mu,\pi_{x^\mu}\lambda^\mu]:=\int_{\tau_1}^{\tau_2}d\tau
\left[\dot{x}^\mu\pi_{x^\mu}-H_c+\lambda^{x^\mu}\phi_{x^\mu}\right].
\end{eqnarray}

By consistency, the primary constraints are evolved in time and from this
it is straightforward to observe that four Lagrange multipliers are fixed,
\begin{eqnarray}\label{mult:3}
\lambda^q=u\frac{p}{m},\qquad\lambda^t=u,\qquad\lambda^{p}=-um\omega^2q,\qquad\lambda^{p_t}=0,
\end{eqnarray}
and that $\phi_u$ generates a secondary constraint
\begin{eqnarray}
\Phi=p_t+\frac{1}{2m}\left(p^2+m^2\omega^2q^2\right)\approx0.
\end{eqnarray}
The time evolution of the secondary constraint yields a relationship among the
Lagrange multipliers, which strongly vanishes after plugging into it the explicit
values of the Lagrange multipliers (\ref{mult:3}), and therefore, no more constraints
arise.

The Poisson brackets among the constraints are computed in order to classify
them. From this, it is straightforward to obtain that $\delta_1:=\phi_u$ and
\begin{eqnarray}
\fl \delta&:=&\gamma+\frac{p}{m}\phi_q+\phi_t-m\omega^2\phi_p\nonumber\\
\fl &=&\pi_t+\frac{p\pi_q}{m}-m\omega^2q\pi_p+
\frac{p}{m}\left(\frac{\partial B}{\partial
q}-\frac{p}{2}\right)+m\omega^2q \left(\frac{q}{2}-\frac{\partial
B}{\partial p}\right)+\frac{\partial B}{\partial t}
\end{eqnarray}
are first-class constraints and that the remaining ones are second-class constraints.

The first class Hamiltonian is $H=u\delta$. Moreover, the extended
action principle takes the form
\begin{eqnarray}
S[x^\mu,\pi_{x^\mu},\lambda^e,\Lambda^\xi]:=\int_{\tau_1}^{\tau_2}d\tau[
\dot{x}^\mu\pi_{x^\mu}-H-\lambda^e\Gamma_e-\Lambda^\xi\chi_\xi].
\end{eqnarray}
where $e=1,2$ y $\xi=1,...,4$ . The last step of the procedure is to
make the counting of degrees of freedom which is
$\frac{1}{2}[2(5)-4-2(2)]=1$, this number agrees with the original
system.

As final comment, we will consider the boundary term which arises
from the gauge transformation of the action principle generated by
the first-class constraints $\delta$. The boundary term which arises
from the gauge transformation generated by $\delta$ is
\begin{eqnarray}
M&=&\varepsilon\left(\frac{\partial\delta}{\partial\pi_\mu}\pi_\mu-\delta\right)\nonumber\\
&=&\varepsilon\left[\frac{p}{m}\left(\frac{p}{2}-\frac{\partial
B}{\partial q}\right)-m\omega^2\left(\frac{q}{2}-\frac{\partial
B}{\partial p}\right)- \frac{\partial B}{\partial t}\right],
\end{eqnarray}
therefore, in the general case, the extended action principle is not fully gauge invariant.
Nevertheless, if the boundary term, $B$, is chosen to be  \cite{HTV1992,MV2002}
\begin{eqnarray}
B=\frac{1}{2}qp,
\end{eqnarray}
the extended action principle becomes fully gauge invariant!

\subsection{A system originally defined by first-class constraints only}
The next example is the system described by the Hamiltonian action
principle
\begin{eqnarray}\label{act1:4}
S[q^i,p_i,u^i]:=\int^{\tau_2}_{\tau_1}d\tau
\left[\dot{q}^ip_i-u^i\gamma_i\right],\qquad i=1,2, \ldots, N,
\end{eqnarray}
where $\gamma_i:=(p_i-q_i) \approx 0$ are $N$ first-class constraints, $u^i$
are Lagrange multipliers, and so the system has zero degrees of freedom.

Next, a boundary term is added to the action (\ref{act1:4})
\begin{eqnarray}
S[q^i,p_i,u^i]=\int_{\tau_1}^{\tau_2}d\tau\left[{ \dot q}^i p_i -
u^i \gamma_i -\frac{d B(q,p)}{d\tau} \right].
\end{eqnarray}
By interpreting the integrand as a Lagrangian function, i.e.,
$\mathcal{L}={\dot q}^i p_i -u^i \gamma_i - \frac{\partial
B}{\partial q^i} {\dot q}^i - \frac{\partial B}{\partial p_i} {\dot
p}_i$, the momenta $\pi_{x^\mu}$ canonically associated to the
coordinates $(x^{\mu})=(q^i,p_i,u^i)$ must be defined from which
$3N$ primary constraints arise
\begin{eqnarray}
\fl \phi_{q^i} :=  \pi_{q^i}-p_i + \frac{\partial B}{\partial q^i}
\approx 0, \quad \phi_{p^i} := \pi_{p^i} + \frac{\partial
B}{\partial p^i} \approx 0, \quad \phi_{u^i} := \pi_{u^i}\approx 0.
\end{eqnarray}
The computation of the canonical Hamiltonian leads to $H_c:=u^i \gamma_i $,
and the total action principle becomes
\begin{eqnarray}
\fl
S[x^\mu,\pi_{x^\mu},\lambda^{x^\mu}]=\int_{\tau_1}^{\tau_2}d\tau\left
[
\dot{x}^\mu\pi_{x^\mu}-H_c-\lambda^{q^i}\phi_{q^i}-\lambda^{p_i}\phi_{p_i}
- \lambda^{u^i} \phi_{u^i} \right ].
\end{eqnarray}
From the evolution of the primary constraints $\phi_{q^i}$ and
$\phi_{p^i}$, the first $2N$ Lagrange multipliers are determined
\begin{eqnarray}\label{mult:4}
\lambda^{q^i}=u^i, \quad \lambda^{p^i}=u^i,
\end{eqnarray}
while from the evolution of $\phi_{u^i}$, the constraints $\gamma_i$ of the
original description of the system (\ref{act1:4}) appear again, this time as
secondary constraints
\begin{eqnarray}
\gamma_i = p_i - q_i \approx 0,
\end{eqnarray}
whose evolution does not generate additional constraints. A straightforward
computation shows that $\phi_{u^i}$ and
\begin{eqnarray}
\delta_i&=&\gamma_i+\phi_{q^i}-\phi_{p^i}\nonumber\\
&=&\pi_{q^i}-\pi_{p^i}-q^i+\frac{\partial B}{\partial
q^i}-\frac{\partial B}{\partial p^i}
\end{eqnarray} are
first-class and that $\phi_{q^i}$ and $\phi_{p_i}$ are second-class.
The first-class canonical Hamiltonian becomes $H=u^i \delta_i$. The
last step is to make the counting of the degrees of freedom of the
system, which is $\frac{1}{2} \left ( 2\times 3N - 2N -2 \times 2N
\right ) =0$, that is in agreement with the original description for
the system given by the action (\ref{act1:4}).

Finally, the boundary term which arises from the gauge
transformation generated by the first-class constraints acquire the
form
\begin{eqnarray}
M_{\delta}=\varepsilon\left(\frac{\partial B}{\partial
p^i}-\frac{\partial B}{\partial q^i}+q^i\right).
\end{eqnarray}
Thus, the extended action principle will be gauge invariant or not
depending on which boundary term $B$ had be chosen.
\subsection{Action for a particle on a sphere $S^2$}

Now is considered the ``free particle" restricted to move through a sphere,
this system contains second-class constraints due to the restriction of the
motion to the sphere. The system is described by the Hamiltonian action
principle \cite{book}
\begin{eqnarray}\label{act-sphe}
S[q^\alpha,p_\alpha,v^\alpha]:=\int_{\tau_1}^{\tau_2}d\tau[\dot{q}^\alpha
p_\alpha-H_0-v^\alpha\varphi_\alpha],\qquad \alpha=1,\ldots,4,
\end{eqnarray}
with $(q^{\alpha})=(x,y,z,u)$,
$H_0=\frac{|\vec{p}|^2}{2m}+\frac{u}{2}(|\vec{q}|^2-R^2)$ is the
first-class Hamiltonian and $R$ is a constant; $\varphi_1:=p_u
\approx 0$, $\varphi_2:=|\vec{q}|^2-R^2 \approx 0$,
$\varphi_3:=\vec{q}\cdot\vec{p} \approx 0$ and
$\varphi_4:=\frac{|\vec{p}|^2}{m}-u  |\vec{q}|^2 \approx 0$ are
second-class constraints and $v^\alpha$ are their respective
Lagrange multipliers. The system has two degrees of freedom. Next, a
time boundary term is added to the action (\ref{act-sphe})
\begin{eqnarray}
S[q^\alpha,p_\alpha,v^\alpha]:=\int^{\tau_1}_{\tau_2}d\tau[\dot{q}^\alpha
p_\alpha-H_0-v^\alpha\varphi_\alpha-\frac{d}{d\tau}B(q^\alpha,p_\alpha)].
\end{eqnarray}

By interpreting the integrand as a Lagrangian function, the momenta
$(\pi_{x^\mu})=(\pi_{q^\alpha},\pi_{p_\alpha},\pi_{v^\alpha})$
canonically conjugated to the coordinates
$x^\mu=(q^\alpha,p^\alpha,v^\alpha)$ must be defined from which
twelve primary constraints arise
\begin{eqnarray}\label{primcon:5}
\fl \begin{array}{cccccc}
\phi_{q^\alpha}:=\pi_{q^\alpha}-p_\alpha+\frac{\partial B}{\partial
q^\alpha}\approx0, &&\phi_{p^\alpha}:=\pi_{p^\alpha}+\frac{\partial
B}{\partial p_\alpha}\approx0,
&&\phi_{v^\alpha}:=\pi_{v^\alpha}\approx0, \\
\end{array}
\end{eqnarray}

The computation of the canonical Hamiltonian leads to
$H_c=\frac{|\vec{p}|^2}{2m}+(\frac{u}{2}+v^2)(|\vec{q}|^2-R^2)+v^1
p_u+ v^3\vec{q}\cdot\vec{p}+v^4(\frac{|\vec{p}|^2}{m}-u
|\vec{x}|^2)$ and the total action principle becomes
\begin{eqnarray}
S[x^\mu,\pi_{x^\mu},\lambda^{^\mu}]:=\int_{\tau_1}^{\tau_2}d\tau[\dot{q}^\mu
\pi_{x^\mu}-H_c-\lambda^{x^\mu}\phi_{x^\mu}].
\end{eqnarray}

By consistency the constraints must be evolved and it is easy to see
that the evolution of $\phi_x, \ldots , \phi_{p_u}$ fix the first eight
Lagrange multipliers
\begin{eqnarray}\label{mult:5}
    \lambda^{q^i} &=& \frac{p^i}{m}(1+2v^4)+q^iv^3, \qquad \lambda^{p^i}=uq^i(2v^4-1)-2q^iv^2-p^iv^3, \nonumber\\
    \lambda^u &=& v^1,\qquad\qquad\quad\quad
    \qquad\lambda^{p_u}=v^4 \mid \vec{x}\mid ^2-\frac{1}{2}(\mid \vec{q}\mid^2-R^2),
\end{eqnarray}
while the time evolution of $\phi_{u^\alpha}$ yield the original
constraints from (\ref{act-sphe}), i.e.,
\begin{eqnarray}\label{seccons:5}
\varphi_1 &=& p_u \approx0,\quad\,\,\,\, \varphi_{2} =|\vec{q}|^2-R^2\approx0, \nonumber\\
\varphi_3 &=& \vec{q}\cdot\vec{p}\approx0, \quad \varphi_4
=\frac{\mid \vec{p}\mid^2}{m}-u|\vec{q}|^2\approx0,
\end{eqnarray}
whose evolution give no more constraints. A straightforward
computation shows that $\phi_{v^\alpha}$ are first-class and that
the remaining ones are second-class. By using these information the
number of physical degrees of freedom of the system is computes
which turns out to be $\frac{1}{2}[2(12)-12-2(4)]=2$, which is in
agreement with the original description for the system given by the
action (\ref{act-sphe}).

Finally, the first-class constraints $\phi_{v^\alpha}$ of the system
are linear and homogeneous therefore the boundary term which arise
from the gauge transformation of the extended action is zero and
hence the system is fully invariant no matter what boundary term had
been chosen.

\subsection{Action defined by a time boundary term only}
This system is defined by an action principle of the form
\begin{eqnarray}\label{bound-only}
S[q^i] &=& B (q^i)\mid^{\tau_2}_{\tau_1}, \quad i=1,\ldots, N.
\end{eqnarray}
By introducing the time boundary term into the integral action leads to
\begin{eqnarray}\label{bound}
S[q^i] &=& \int^{\tau_2}_{\tau_1}d\tau \frac{d B}{d \tau}  =
\int^{\tau_2}_{\tau_1}d \tau \left( \frac{\partial B}{\partial q^i}
{\dot q}^i \right ) .
\end{eqnarray}

The next step is to interpret the action (\ref{bound}) as a
Lagrangian one, i.e., the $q$'s are interpreted as configuration
variables and so Dirac's method calls for the definition of their
canonically conjugate momenta $\pi_{q^i}$. Thus, from the definition
of the momenta and the Lagrangian $\mathcal{L}= \frac{\partial
B}{\partial q^i} {\dot q}^i$, the primary constraints

\begin{eqnarray}\label{pbound}
\phi_i &:=& \pi_{q^i} - \frac{\partial B}{\partial q^i} \approx 0.
\end{eqnarray}
arise. The canonical Hamiltonian $H_c = {\dot q}^i \pi_{q^i} -
\mathcal{L}$ identically vanishes. On the other hand, the evolution
of the primary constraints is strongly zero and so (\ref{pbound})
are first-class. The Hamiltonian action principle acquires the form

\begin{eqnarray}
S[q^i,\pi_{q^i}, u^i]&=&\int^{\tau_2}_{\tau_1}d\tau\left({\dot q}^i
\pi_{q^i} - \lambda^i\phi_i\right).
\end{eqnarray}

The dynamics of this theory is pure gauge in the sense that the number of
physical degrees of freedom is zero, $\frac12\left(2N- 2N\right)=0$, and
evolution in $\tau$ is the unfolding of the gauge symmetry.

By other hand, the boundary term which come from the gauge
transformations generated by the first-class constraints is the
following
\begin{eqnarray}
M_\phi=\varepsilon\frac{\partial B}{\partial q^i}.
\end{eqnarray}

By one hand, this example is relevant because it illustrates the
role played by boundary terms only. By other hand, this result can
be expressed as follow

\begin{theorem}
Let $\mathcal{M}$ be a $m$-dimensional manifold with boundary
$\partial\mathcal{M}$ of dimension $m-1$. Let $L_B(q^i,\dot{q}^i)$
and $L_F(q^i)$ be two Lagrangian functions defined on $\mathcal{M}$
and $\partial \mathcal{M}$ respectively, where $q^i$ are $N$
coordinates which label the points of the configuration space.
Therefore, if $L_B$ such as
\begin{eqnarray}
L_B(q^i,\dot{q}^i)=\frac{d}{d t}L_F(q^i)
\end{eqnarray}
then the theory defined by $L_B$ is topological.
\end{theorem}

The theorem is still valid if $L_F$ depend on the momenta $p^i$ or
the higher $n$-th derivative of the coordinates\footnote{The
demonstration of the theorem when $L_F$ depends on the higher
derivatives of the coordinates do not has been deployed here because
it is out of the spirit of the paper.} $\frac{d^nq}{d\tau^n}$,
namely, if the Lagrangian $L_B$ can be expressed as the total
derivative of some function $L_F$ the number of the degrees of
freedom will be zero, namely the theory always will be topological.
This theorem is true for system with finite numbers of degrees of
freedom and as well as for field theories as it is seen in the
appendix and in \cite{gm}.


\section{Concluding remarks}

\endnumparts

It has been studied how to manage time boundary terms in the theoretical
framework of Dirac's canonical analysis. The strategy consist in the
introduction of the time boundary term into the action principle, thus
enlarging the original set of configuration variables. The resulting action is
interpreted as a Lagrangian one to which the canonical analysis can be
applied. In this approach the time boundary conditions of the new action
principle are on the new full set of configuration variables, as usual. The
information of the boundary term is encoded in the new second-class
constraints as well as in the redefinition of the new first-class constraints.

The approach followed here, can also be applied to the case when the original
action principle is endowed with arbitrary symplectic structures instead of
canonical ones \cite{cmv}.

It is worthwhile to mention that the boundary term added can also include the
original Lagrange multipliers. The analysis in such a case can be carried out
following essentially the same steps made in the current paper.

As final comments, the approach developed here can also be applied
to mini-superspace models (when the space-time has specific
symmetries), such as cosmological models [see, for instance,  Ref.
[13-19]. Moreover, these way to deal with boundary term could be
applied into the zero-Hamiltonian problem in 2D gravity
\cite{delfrate,const} and into topological field theories \cite{gm}.
Finally, the approach developed here can extended to its complex
counterpart and analyze complex canonical transformations
\cite{dector}.

\ack This work was presented in the parallel session of the 12th
Marcel Grossmann Meeting held in Paris France, 2009. We thank M
Montesinos and JD Vergara for very fruitful discussions on the
subject. This work was supported in part by CONACYT, Mexico, Grant
No. 56159-F.


\appendix

\section{The Stokes's theorem as a topological field theory}

In this appendix we will consider the Stokes theorem in two
dimensions as special case of the theorem 1 for field theory. The
Stokes theorem can be read as
\begin{eqnarray}\label{st}
\int_{{\mathscr{M}}^n}d\omega=\int_{\partial{\mathscr{M}}^n}\omega,
\end{eqnarray}
where ${{\mathscr{M}}^n}$ is a $n$-manifold with boundary, $\omega$
is an $(n-1)$-form. For the sake of simplicity, let us consider a
$2$-manifold ${{\mathscr{M}}^2}$. Let $(x,y)$ be local coordinates
that label the points of ${{\mathscr{M}}^2}$. Therefore,
$\omega=X(x,y)dx+Y(x,y)dy$. By one hand, the Stokes theorem
(\ref{st}) seen as an action principle acquires the form
\begin{eqnarray}\label{act-st}
S[X,Y]:=\alpha\int_{{\mathscr{M}}^2}\left[\partial_xY(x,y)-\partial_yX(x,y)\right],
\end{eqnarray}
where $\alpha$ is the constant of proportionality which absorbs the
unities. By other hand, we consider the parametrization of the
coordinates $(x,y)$
\begin{eqnarray}
x=x(\tau,\sigma),\quad\mbox{and}\quad y=y(\tau,\sigma).
\end{eqnarray}
The range of the parameters are $\tau_1\leq\tau\leq\tau_2$ and
$\sigma_1\leq\sigma\leq\sigma_2$, namely, we parameterize the
surface ${{\mathscr{M}}^2}$ with a square. Solving the components of
(\ref{act-st}) in terms of the parameters $(\tau,\sigma)$, we obtain
\begin{eqnarray}\label{act-h-st}
S[X,Y]:=\alpha\int_{{\mathscr{M}}^2}d\tau\wedge d\sigma
\left(\dot{X}x'+\dot{Y}y'-\dot{y}Y'-\dot{x}X'\right).
\end{eqnarray}
where the dot and the apostrophe denote derivation with respect to
$\tau$ and $\sigma$ respectively.

In order to make the counting of degrees of freedom, the momenta
$(\pi_i)=(\pi_X,\pi_Y,p_x,p_y)$ canonically associated to the
coordinates $(q^i)=(X,Y,x,y)$ must be defined, and these arise four
primary constraints
\begin{eqnarray}\label{const-st}
\phi_X:=\pi_X-\alpha x'\approx0,\quad \phi_Y:=\pi_Y-\alpha y'\approx0,\nonumber\\
\phi_x:=p_x+\alpha X'\approx0,\quad\phi_x:=p_y+\alpha Y'\approx0.
\end{eqnarray}
A straightforward computation implies that the canonical Hamiltonian
vanishes, and so the action principle acquires the form
\begin{eqnarray}
S[q^i,\pi_i,\lambda^i] =\int_{\tau_1}^{\tau_2}d\tau
\int_{\sigma_1}^{\sigma_2}d\sigma\,\left({\dot q}^ip_i-\lambda^i
\phi_i\right)\quad i=1,...,4.
\end{eqnarray}
By consistency, the primary constraints (\ref{const-st}) must be
evolved respect to the parameter $\tau$. These $\tau$-evolution are
strongly zero and therefore there are no more constraints. Moreover,
the algebra of constraints tell us that the constraints
(\ref{const-st}) are first-class.

The extended phase space is parameterized by $4$ configuration
variables $q^i$ and the corresponding $4$ canonical momenta $\pi_i$,
there are $4$ first-class and $0$ second-class constraints.
Therefore the system has $\frac12(2\times4-2\times2)=0$ physical
degree of freedom per point of $\sigma$, namely, the theory defined
by the Stokes theorem is topological as was pointed out in the
theorem 1.

As final comment it is worth noticing that besides the left side of
the Stokes theorem (\ref{st}) is topological, the canonical analysis
of the right side of (\ref{st}) revels that the theory defined by
$L_F=\alpha\omega$ has one degree of freedom \cite{gallardo}.
Namely, the theory described by (\ref{act-st}) is other example of
the theories defined in a manifold with boundary which are
topological in the bulk and has local degrees of freedom into its
boundary \cite{gm}.


\end{document}